# Using Fuzzy Logic to Evaluate Normalization Completeness for An Improved Database Design

M. Rizwan Jameel Qureshi [1]
*Faculty of Computing & Information Technology of King Abdul Aziz University, Jeddah, Saudi Arabia*
*Email: anriz@hotmail.com*

Mahaboob Sharief Shaik [2]
*Faculty of Computing & Information Technology of King Abdul Aziz University, Jeddah, Saudi Arabia*
*Email: mshaik@kau.edu.sa*

Nayyar Iqbal [3]
*Department of Computer Science of Agricultural University, Faisalabad, Pakistan*
Email: nayyar_ciit@yahoo.com

***Abstract*—A new approach, to measure normalization completeness for conceptual model, is introduced using quantitative fuzzy functionality in this paper. We measure the normalization completeness of the conceptual model in two steps. In the first step, different normalization techniques are analyzed up to Boyce Codd Normal Form (BCNF) to find the current normal form of the relation. In the second step, fuzzy membership values are used to scale the normal form between 0 and 1. Case studies to explain schema transformation rules and measurements. Normalization completeness is measured by considering completeness attributes, preventing attributes of the functional dependencies and total number of attributes such as if the functional dependency is non-preventing then the attributes of that functional dependency are completeness attributes. The attributes of functional dependency which prevent to go to the next normal form are called preventing attributes.**

***Index Terms*—Normalization completeness, Conceptual model, Relation, Functional dependency, Total attributes, Completeness attributes, Preventing attributes**

## 1. Introduction

Conceptual model describes a complete framework for a database to be designed. Conceptual model is represented by the entity-relationship diagram or entity-relationship model that includes entities, their attributes and relationships between them [1][2]. We measure the normalization completeness for conceptual model using quantitative fuzzy functionality in two steps. Initially, we are finding the normal form of the relation by analyzing different normalization techniques up to BCNF such as checking composite attributes, partial dependencies and transitive dependencies of the relation. Normalization process [3] requires a set of dependencies to be defines for every problem. Further, we are using fuzzy membership values to scale normal form of the relation between 0 and 1.

The introduced normalization completeness determines how much the normal form is closer to the next normal form. The quality model of ISO 9126 defines functionality as a collection of attributes that engage on the existence of a set of functions and their specific properties. The functions are that satisfy stated needs which are as follow suitability, accuracy, interoperability, compliance and security [4].

The remainder of this paper is organized as follows: Section 2 describes related work. Section 3 defines the problem statement. Section 4 presents the hypothesis. Section 5 illustrates the evaluation as a proof of concept using a case study. Conclusion and future work are given in the final section.

## 2. Related Work

Lindlan et al. are the first to articulate a systematic framework to help understanding quality in the context of conceptual modeling [5, 6]. Previous attempts merely resulted in lists of unstructured, imprecise and often overlapping quality properties. Lindlan et al. framework is the only (framework) that contained both theory and empirical validation. The semantic quality of conceptual modeling script is difficult to evaluate directly as it is hard (and perhaps even impossible) to know reality, externalize this knowledge (which would mean building another script) and agree upon it. When evaluating semantic quality, users can only refer to their perception of reality, which is obtained through observation and internalization. The questions to find out which filter to put upon reality by our observation possibly depends on many factors such as previously acquired knowledge, perceptual psychology effects, cognitive abilities, and ontological and epistemological standpoints taken [5,6].

The research in [7] describes how to measure the semantic quality of the conceptual model using completeness. The method used to measure quantitative completeness first checks the functional dependencies. Transformation rules are applied to conceptual model and convert it into multi-graph. The concept of membership values and fuzzy hedging is used. The completeness measurement identifies the effort required for the conceptual model to transform into another conceptual model in the improved form. The quality of the conceptual model is measured using new introduced fuzzy completeness index (FCI) [8]. By considering the functional dependencies of the conceptual model the completeness of the conceptual model is measured quantitatively. The functional dependencies of the conceptual model are mapped on the TAS graph, and the completeness of conceptual model is measured by using new FCI approach [8]. The value of the FCI determines the completeness of the conceptual model.

On schema transformation, the paper [9] presents new definitions for 'primary key', 'non-key attributes', 'key attributes' and 'functional dependency'. Schema transformation rules are also proposed in [9]. Two quality metrics are introduced namely normalization index and completeness index. Rules are applied on case studies of conceptual model to measure the normalization index and completeness index [9]. Structural complexity of a conceptual model is measured with two parameters namely modifiability and understandability [10]. Modifiability of the conceptual model is measured with effort to change. Understandability of the conceptual model is discussed by correctness and its main types are syntactic and semantic [10].

The quality of a conceptual model is divided into three types 1) syntax 2) semantic 3) pragmatic [11]. Hussain et al. [11] introduced an approach of schema transformation to improve the semantic quality of the conceptual model. The rules depend on the functional dependencies given for the conceptual model. The normal form of the conceptual model is measured up to BCNF using multiple case studies [11]. Hussain et al. [3] [12] described the eliminating process of normalization. The violations while performing normalization prevent the designer to go to next form after BCNF. The normalization algorithms depend upon inclusion, multi-valued, functional and join dependencies. Removing these dependencies from given problem is a time consuming and difficult task [3] [12].

The paper [11] describes the effort based completeness index for entity relationship diagram by considering the satisfying index and the effort to change for a functional dependency. The comparison of completeness index, fuzzy completeness index and effort based completeness index on different conceptual models are also shown in [13]. Two different conceptual models of the same problem can have same completeness index but there effort based completeness index will have different value [13].

Thalheim [14] recommends various design quality parameters for conceptual model such as flexibility, naturalness and completeness. He [14] further defines that completeness is the representation of all relevant features of the application domain. The relations are first normalized in order to obtain fuzzy relational database [15]. Fuzzy database relation has many advantages over standards database. Standard normalization depends upon the functional dependency therefore fuzzy functional dependency must be defined for fuzzy relational database normalization [15]. A model is proposed in [16] for normalization of database and functional dependencies are also introduced into the systems that needs to be considered. Inclusion dependencies are also introduced into the system [16]. The proposed model of [16] can be used to perform normalization of large database systems to remove functional dependencies.

A concept of functional dependency is introduced as rough set and relation database [17]. Functional dependency discovering algorithm is divided into two parts. In the first part, a hypothesis is defined regarding functional dependency to authenticate it against relation. In the second part, hypothesis validation is done by checking it row by row [17]. XML tree, path expressions and DTD and XML functional dependencies are described in [18] to DTD base relational schema mapping algorithm so that the semantic and structure of key can be preserved. Markus [19] describes the role of insertion, deletion and update anomalies. Semantic dependencies are the base of the code normal forms. Semantic dependencies define only one function whereas functional dependencies relates to group of attributes [19].

The normalization theory was proposed by E. F. Codd in 1970's, and the rough set theory was introduced by Pawlak in 1982 [20]. While considering only the functional dependency then BCNF is the highest normal form in a relational database. In the relational normalization theory, functional dependency and normal form perform the function of a kernel [20]. An automatic database normalization approach is introduced in [21]. Three structures are proposed to represent functional dependencies of the relational database that are dependency graph, direct graph matrix and dependency matrix [21]. Functional dependencies of the relation are represented by dependency graph diagram in which composite key is above the dotted line whereas other attributes of the functional dependencies are below the dotted line [21].

A concept of normal form for XML documents is introduced in [22] so that redundancy and update anomalies can be controlled. Further, a comparison is made between XML normal form, BCNF and nested normal form. A hierarchical schema in [23] represents XML database schema and corresponding normal forms (first normal form (1NF) and second normal form (2NF)) for XML database schema. It [23] also presents the algorithm eliminating redundant schemas and normalization design algorithm up to 2NF. A concept, of functional independent normal form, is introduced in [24] considering the 'functional dependencies attributes' on the left hand side and commonly known 'determinant and attributes' on the right hand side of the functional dependency. The normal form of the database relation must be in BCNF and the conditions between the attributes of the functional dependencies must be present as follows [24].

$$A \rightarrow B \text{ or } B \rightarrow A \text{ or } A >< B$$

Fuzzy logic [25][26] determines the membership values in numerical form that are '0' and '1'. Zero '0' means no membership and one '1' means complete membership in the condition as follows.

$$0 < x < 1$$

## 3. Problem Statement and Hypothesis

The problem statement is described as follows.
How to measure the completeness of normal form to determine that how much it is closer to the next normal form?

In the hypothesis we measure normalization completeness for the conceptual model using quantitative fuzzy functionality up to BCNF. In this research three hypothesis are considered as follows.

H1: No membership value.
H2: Partial membership value.
H3: Complete membership value.

## 4. Experiment and Analysis

Normalization completeness determines how much the normal form is closer to the next normal form. We measure normalization completeness up to BCNF.

NC= N + Fuzzy Functionality of the conceptual model ………… (1)

NC = N + (((completeness attributes of the FD's / total attributes) + (1-(preventing attributes of the FD's/ total attributes)))…. (2)

NC is normalization completeness and N is current normal form determined by analyzing different normalization techniques discussed in [1].

**4.1 Proof**

In this we prove the normalization completeness for conceptual model. Fuzzy sets defined by Lotif Zadeh is given by

M: x → [0, 0.01……. 0.99, 1]

or

M: x → [no membership value, partial membership value, complete membership value]

where M is fuzzy set and x describes the membership value. In which no membership value = 0, 0 < partial membership value < 1 and complete membership value = 1.

Fuzzy sets value ranges from 0 to 1.

Therefore

$$0 \le x \le 1 \ldots\ldots\ldots\ldots\ldots\ldots (3)$$

suppose

x = (((completeness attributes of the FD's / total attributes) + (1-(preventing attributes of the FD's / total attributes))) / 2)

substituting the value of x in (3)

0 ≤ (((completeness attributes of the FD's / total attributes) + (1-(preventing attributes of the FD's / total attributes))) / 2) ≤ 1

Consider the total attributes of the functional dependency is n

**Complete Membership Value**
if completeness attributes = total attributes
then preventing attributes = 0
therefore

$x = (((n / n) + (1 - (0 / n))) / 2) = (((n / n) + (1 - 0)) / 2) = ((1 + 1) / 2) = (2 / 2) = 1$

hence it proves the complete membership is equal to 1

**No Membership Value**
if preventing attributes = total attributes
then completeness attributes = 0
therefore

$x = (((0 / n) + (1 - (n / n))) / 2) = (((0 - 1) + (1 - 1)) / 2) = ((0 + 0) / 2) = (0 / 2) = 0$

hence it proves the no membership is equal to 0

**Partial Membership Value**
if completeness attributes ≤ total attributes & preventing attributes ≤ total attributes then x = partial membership value.

Hence it is proved that completeness attributes makes the normal form closer to the next normal form whereas preventing attributes decreases the completeness from the next normal form.

**Process for Finding the Normal Form**

Following is the process for finding the normal form in which we analyze different normal forms techniques up to BCNF, the obtained value is assigned to N.

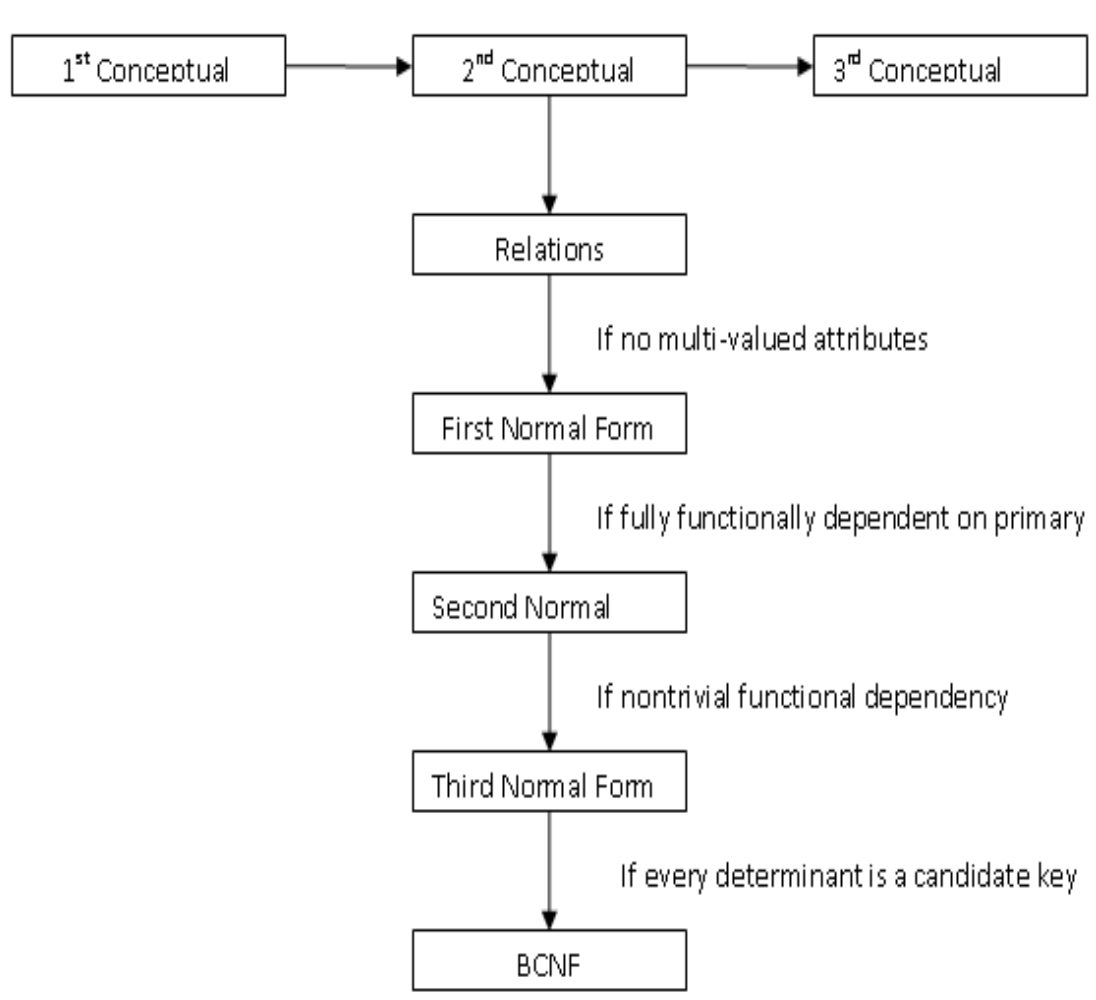

Fig. 1: Process for Finding the Normal Form

**4.2 Case Study**
The details of case study are discussed in [27].

FD1: propertyNo, iDate → iTime
FD2: propertyNo, iDate → comments
FD3: propertyNo, iDate → staffNo
FD4: propertyNo, iDate → sName
FD5: propertyNo, iDate → carReg
FD6: propertyNo → pAddress
FD7: staffNo → sName
FD8: staffNo, iDate → carReg
FD9: carReg, iDate, iTime → propertyNo
FD10: carReg, iDate, iTime → pAddress
FD11: carReg, iDate, iTime → comments
FD12: carReg, iDate, iTime → staffNo
FD13: carReg, iDate, iTime → sName
FD14: staffNo, iDate, iTime → propertyNo
FD15: staffNo, iDate, iTime → pAddress
FD16: staffNo, iDate, iTime → comments

Relation of StaffPropertyIspection are StaffPropertyIspection(propertyNo, iDate, iTime, pAddress, comments, staffNo, sName, carReg) and current normal form is 1 that means N=1.

We now find the normalization completeness: non-preventing functional dependencies are FD1, FD2, FD3, FD4, FD5, FD9 FD10, FD11, FD12, FD13, FD14, FD15 and FD16 where as preventing functional dependencies are FD6, FD7 and FD8. Completeness attributes of the FD's is '8', preventing attributes of the FD's are '6' and total attributes are equal to '8'.

NC= N + Fuzzy Functionality of the conceptual model
NC= N + (((completeness attributes of the FD's / total attributes) + (1- (preventing attributes of the FD's / total attributes))) / 2).

$$\begin{aligned}NC &= 1 + (((8 / 8) + (1 - (6 / 8))) / 2) \\ &= 1 + (((1) + (1 - (0.75))) / 2) \\ &= 1 + (((1) + (0.25)) / 2) \\ &= 1 + (1.25 / 2) \\ &= 1 + 0.62 \\ &= 1.62\end{aligned}$$

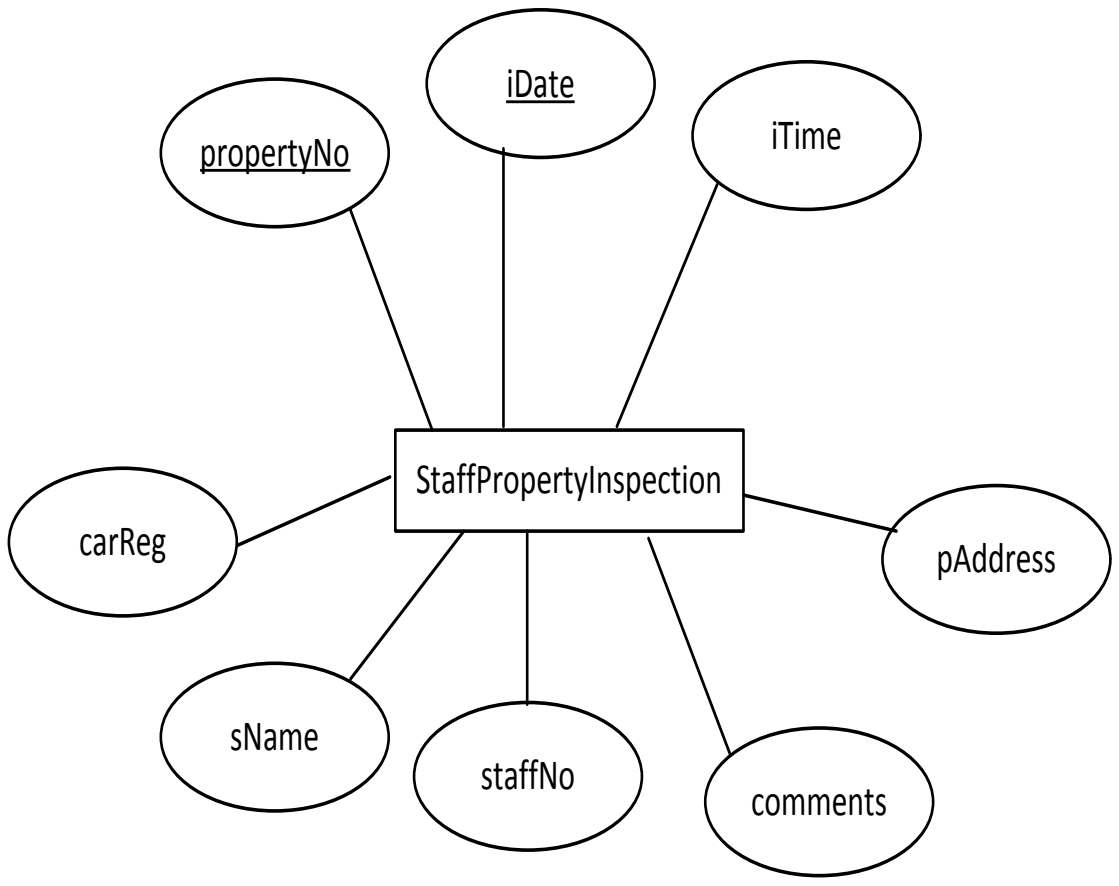

Fig. 2: Conceptual Model

Transformation of figure 2 according to [25]. According to figure 3: StaffInspection(propertryNo, iDate, iTime, comments, staffNo, sName, carReg) and current normal form is 2 therefore N=2. The normalization completeness is non-preventing functional dependencies are FD1, FD2, FD3, FD4, FD5, FD9, FD11, FD12, FD13, FD14 and FD16 where as preventing functional dependencies are FD7 and FD8. Completeness attributes of the FD's are equal to '7' and preventing attributes of the FD's are '4' and total attributes are equal to '7'.

*NC= N + Fuzzy Functionality of the conceptual model*
*= N + (((completeness attributes of the FD's / total attributes) + (1- (preventing attributes of the FD's / total attributes))) / 2)*

= 2 + (((7 / 7) + (1- (4 / 7))) / 2)
= 2 + (((1) + (1- (0.57))) / 2)
= 2 + (((1) + (0.43)) / 2)
= 2 + (1.43 / 2)
= 2 + 0.71
= 2.71

Relation of Property according to figure 3 is: (propertyNo, pAddress) and current normal form is BCNF. Relations according to figure 4 is: Inspection-(propertryNo, iDate, iTime, comments staffNo, carReg) and current normal form is 3 therefore N=3. We now find the normalization completeness: non-preventing functional dependencies are FD1, FD2, FD3, FD4, FD9, FD11, FD12, FD14, and FD16 where as Preventing functional dependency is FD8. Completeness attributes of the FD's are '6', preventing attributes of the FD's are '3' and total attributes are '6'.

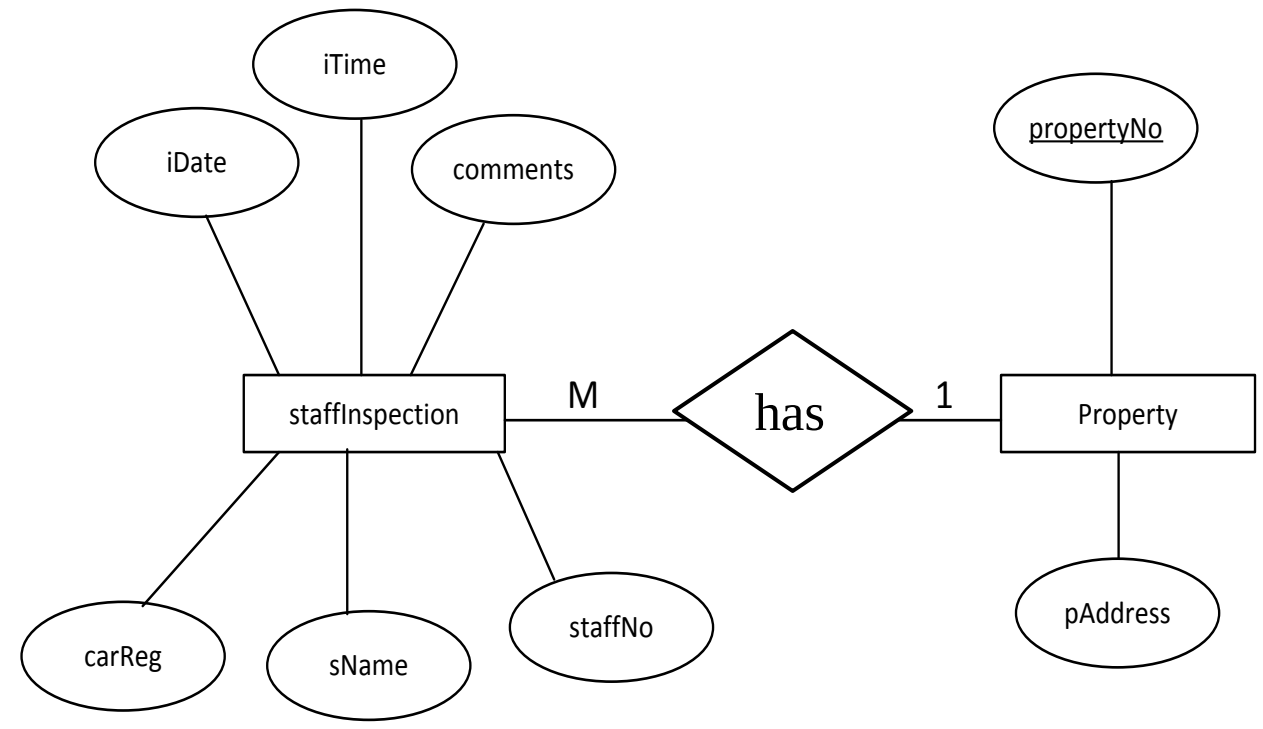

Fig. 3: Improved Conceptual Model after First Transformation

*NC= N + Fuzzy Functionality of the conceptual model*
*NC= N + (((completeness attributes of the FD's / total attributes) + (1- (preventing attributes of the FD's / total attributes))) / 2)*

= 3 + (((6 / 6) + (1- (3 / 6))) / 2)
= 3 + (((1) + (1- (0.5))) / 2)
= 3 + (((1) + (0.5)) / 2)
= 3 + ((1.5) / 2)
= 3 + (1.5 / 2)
= 3 + 0.75
= 3.75

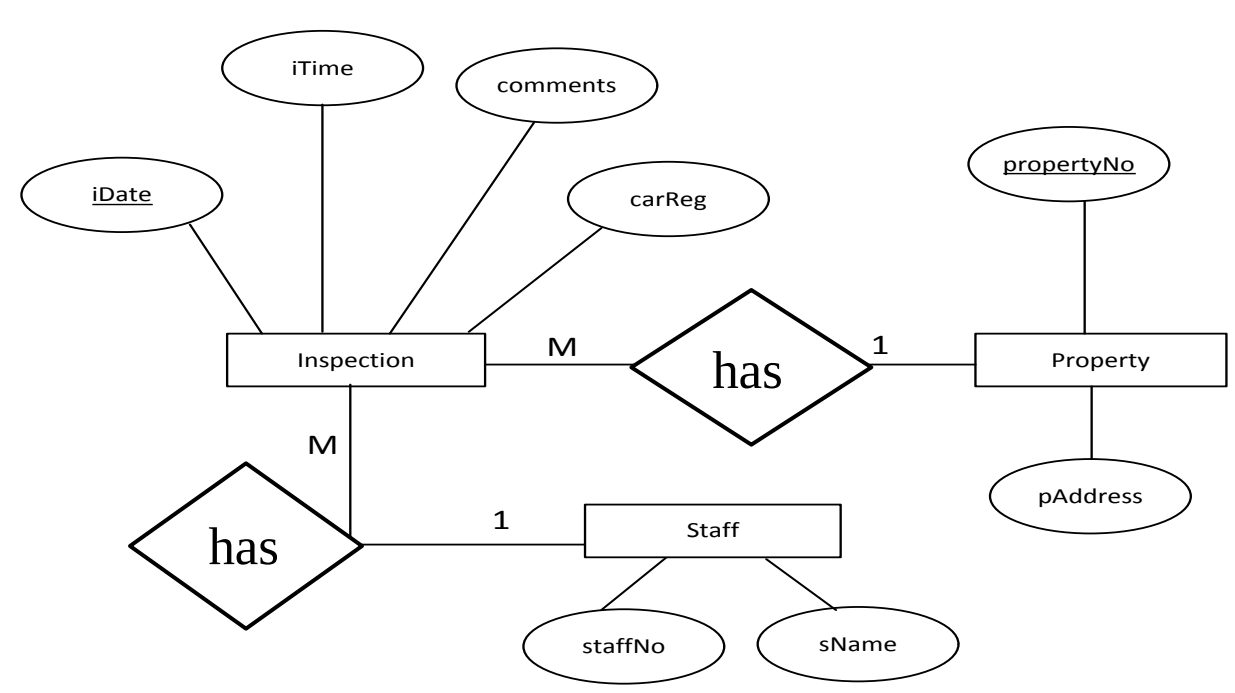

Fig. 4: Improved Conceptual Model after Second Transformation

Relation of Property according to figure 4 is Property (propertyNo, pAddress) and current normal form is BCNF. Relation of Staff [25] according to figure 4 is Staff(staffNo, sName) and current normal form is BCNF.

Table 1: Results of Case Study

| Normalization completeness |
|---|

| Initial Schema | After First Transformation | After Second Transformation |
|---|---|---|
| 1.62 | 2.71 + 4 = 6.71 | 3.75 + 4 + 4 = 11.75 |

## 5. Conclusion

The adapted approach measures normal form for conceptual model up to Boyce Codd Normal Form (BCNF). In this paper, normalization completeness (NC) is measured in two steps. At the first step, value of N (where N stands for normal norm) is determined by analyzing the normal form conditions. In the second step, fuzzy functionality of the conceptual model is determined that is based on hypothesis to determine the completeness of the normal form (Completeness attributes, preventing attributes of functional dependencies and total attributes are considered).

Mathematical proof for the completeness issue is based on three conditions. In first condition, if completeness attributes are equal to total attributes then preventing attributes are equal to zero. This proves that completeness membership is equal to one. In second condition, if preventing attributes are equal to total attributes then completeness attributes are zero. That means no membership is equal to zero. In third condition, if completeness attributes are less than or equal to total attributes and preventing attributes are less than or equal to total attributes, then the resultant value is the partial membership.

The normalization completeness is studied in the context of establishing a conceptual model. The study deals with two or three conceptual models. The model is improved by using transformation rules as per the literature. The model is converted to relational model and it determines normalization completeness for the conceptual model.

**M. Rizwan Jameel Qureshi:** Assistant Professor of Faculty of Computing & Information Technology in King Abdul Aziz University, Saudi Arabia interested in software engineering and database systems.

**Mahaboob Sharief Shaik:** Lecturer, Faculty of Computing & Information Technology in King Abdul Aziz University, Saudi Arabia

**Nayyar Iqbal:** Lecturer at computer science Department at Agricultural University of Faisalabad, Pakistan major database systems and software engineering.